\documentclass[aps,twocolumn,showpacs,prl,preprintnumbers,amsmath,amssymb]{revtex4}
\def\etal{{\it et\ al.}}

\newcommand{\lsim} 
 {\ \raise.35ex\hbox{$<$}\kern-0.75em\lower.5ex\hbox{$\sim$}\ }
\newcommand{\gsim}
 {\ \raise.35ex\hbox{$>$}\kern-0.75em\lower.5ex\hbox{$\sim$}\ }

\newcommand{\mean}[1]{\left<#1\right>}
\newcommand{\means}[1]{\langle#1\rangle}

\def\journal #1#2#3#4{#1 {\bf #2}, #3 (#4)}

\def\PRB{Phys.\ Rev.\ B}
\def\PRL{Phys.\ Rev.\ Lett.}

\def\JPSJ{J.\ Phys.\ Soc.\ Jpn.}

\usepackage{graphicx}
\usepackage{dcolumn}
\usepackage{bm}
\usepackage{amsmath}
\usepackage{times}
\usepackage[usenames]{color}

\usepackage{ulem}

\begin{document}
\title{
Vaporization of Kitaev spin liquids
}
\author{Joji Nasu,$^{1}$ Masafumi Udagawa,$^{2}$ and Yukitoshi Motome$^{2}$} 
 \affiliation{$^{1}$Department of Physics, Tokyo Institute of Technology, Ookayama, 2-12-1, Meguro, Tokyo 152-8551, Japan,\\
$^{2}$Department of Applied Physics, University of Tokyo, Hongo, 7-3-1, Bunkyo, Tokyo 113-8656, Japan}
\date{\today}
\begin{abstract}
 Quantum spin liquid is an exotic quantum state of matter in magnets. This state is a spin analogue of the liquid helium which does not solidify down to the lowest temperature due to strong quantum fluctuations. In conventional fluids, liquid and gas possess the same symmetry and adiabatically connect to each other by bypassing the critical end point. 
We find that the situation is qualitatively different in quantum spin liquids realizing in a three-dimensional Kitaev model; 
both gapless and gapped quantum spin liquid phases at low temperatures are always distinguished from the high-temperature paramagnet (spin gas) by a  phase transition. The results challenge common belief that the absence of thermodynamic singularity down to the lowest temperature is a symptom of a quantum spin liquid. 
\end{abstract}



\maketitle



%
%

%




A magnetic state called quantum spin liquid (QSL), where long-range ordering is suppressed by quantum fluctuations, is a new state of matter in condensed matter physics~\cite{Balents2010}.
Tremendous efforts have been devoted to the realization of QSL, and several candidates were recently discovered in quasi two-dimensional (2D) and three-dimensional (3D) compounds~\cite{Shimizu03,Nakatsuji05,Helton07,Okamoto07,Yamashita10}.
In these compounds, QSL is usually identified by the absence of anomalies in the temperature ($T$) dependence of physical quantities. Namely, it is implicitly supposed that a spin ``gas'' corresponding to the high-$T$ paramagnet is adiabatically connected with QSL. 
This common belief lends itself to the fact that liquid and gas are adiabatically connected with each other in conventional fluids.
In fact, the concept of QSL was originally introduced on the analogy of helium in which the liquid phase is retained down to the lowest $T$ due to strong quantum fluctuations~\cite{Anderson1973}.

In general, however, liquid and gas are distinguished by a discontinuous phase transition, while the adiabatic connection between them is guaranteed beyond the critical end point. 
Hence, a phase transition separating paramagnet and QSL is also expected. 
Nevertheless, the theory for thermodynamics of QSLs has not been seriously investigated thus far, and a thermodynamic phase transition for QSL has not ever been reported beyond the mean-field approximation. 
It is highly nontrivial whether a liquid-gas transition exists in quantum spin systems in a similar manner to that in conventional fluids.
The issue is critical not only for theoretical understanding of QSLs but also for the interpretation of existing and forthcoming experiments.

The lack of theoretical investigation of thermodynamics of QSLs is mainly due to the following two difficulties. 
One is the scarcity of well-identified QSLs. 
It is hard to characterize QSL because spatial quantum entanglement and many-body effects are essential for realizing QSL~\cite{Yan2011,Jiang2012}. 
The other difficulty lies in less choice of effective theoretical tools. 
Any biased approximation might be harmful for taking into account strong quantum and thermal fluctuations.

In this Letter, we solve these difficulties by investigating a 3D extension of the Kitaev model~\cite{Kitaev06}, which supports well-identified QSLs as the exact ground states~\cite{Mandal2009} by applying an unbiased quantum Monte Carlo (MC) simulation without negative sign problem.
By clarifying the phase diagram in the whole parameter space, we show that both the gapped and gapless quantum spin liquid phases exhibit a finite-temperature phase transition to the high-temperature paramagnet.
The results unveil that the ``vaporization'' of the quantum spin liquids are quantitatively different from the conventional liquid-gas transition.

\begin{figure}[t]
\begin{center}
\includegraphics[width=0.9\columnwidth,clip]{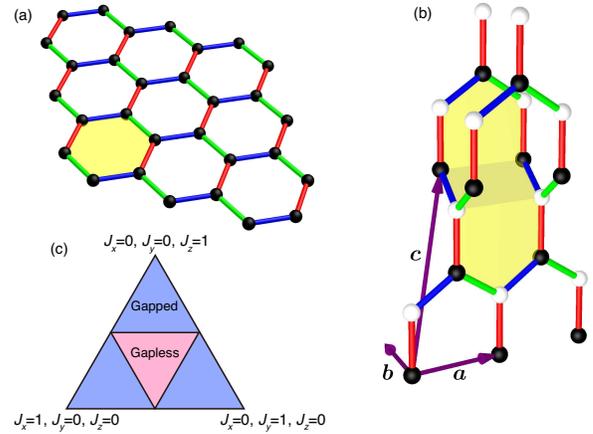}
\caption{(color online).
(a) Two-dimensional honeycomb lattice and (b) three-dimensional hyperhoneycomb lattice. 
Blue, green, and red bonds denote the exchange couplings $J_x$, $J_y$, and $J_z$ in the Kitaev Hamiltonian, respectively.
The shaded plaquette on each lattice represents the shortest loop $p$ for which the $Z_2$ variable $W_p$ is defined.
$\bm{a}$, $\bm{b}$, and $\bm{c}$ represent the primitive translation vectors.
(c) Phase diagram of the Kitaev model at zero temperature, common to the models on the honeycomb and on the 
hyperhoneycomb lattices.
This diagram is depicted on the plane where the condition $J_x+J_y+J_z=1$ is satisfied. There are two kinds of phases, gapped and gapless spin liquids distinguished by the excitation gap.
}
\label{fig1}
\end{center}
\end{figure}

The Kitaev model is a quantum spin model with anisotropic exchange interactions for nearest neighbor spins, whose Hamiltonian is given by 
\begin{align}
{\cal H}=-J_x\sum_{\langle ij\rangle_x}\sigma_i^x\sigma_j^x-J_y\sum_{\langle ij\rangle_y}\sigma_i^y\sigma_j^y-J_z\sum_{\langle ij\rangle_z}\sigma_i^z\sigma_j^z.\label{eq:5}
\end{align}
Here, $\sigma_i^x$, $\sigma_i^y$, and $\sigma_i^z$ are Pauli matrices describing a spin-1/2 state at a site $i$;
$J_x$, $J_y$, and $J_z$ are exchange constants~\cite{Kitaev06}. This model was originally introduced on a honeycomb lattice shown in Fig.~\ref{fig1}(a). 
The interactions $J_x$, $J_y$, and $J_z$ are defined on three different types of the nearest neighbor bonds, $x$ (blue), $y$ (green), and $z$ bonds (red), respectively [see Fig.~\ref{fig1}(a)]. 
This model is exactly solvable by introducing Majorana fermions~\cite{Kitaev06}. 
The ground state of the Kitaev model is a QSL, where spin-spin correlations vanish except for nearest neighbors~\cite{Baskaran2007}.
The ground state phase diagram consists of gapless and gapped QSL phases~\cite{Kitaev06}, as shown in Fig.~\ref{fig1}(c).
The QSL with gapless excitation is stabilized in the center triangle including the isotropic case $J_x=J_y=J_z$, while the QSL with an excitation gap appears in the outer three triangles with anisotropic interactions.
The model has been studied not only from the mathematical virtue of the exact solvability but also from the experimental relevance to some Ir oxides~\cite{Jackeli2009}.

A 3D extension of the Kitaev model is defined on the hyperhoneycomb lattice shown in Fig.~\ref{fig1}(b)~\cite{Mandal2009}. 
This model has relevance to recently-discovered iridates Li$_2$IrO$_3$~\cite{Modic2014,Takayama2014}.
There are three types of nearest neighbor bonds in this lattice as in the honeycomb lattice. 
Many fundamental aspects in the 3D Kitaev model are inherited from the original 2D one, including the exact solvability. 
In particular, the ground state phase diagram is completely the same as that in 2D in Fig.~\ref{fig1}(c)~\cite{Mandal2009}. 
On the other hand, the difference in the spatial dimension may matter to finite-$T$ properties;
while no phase transition is expected at a finite $T$ for the 2D Kitaev model~\cite{Castelnovo2007,Nussinov2008,SM}, we may anticipate a finite-$T$ phase transition in the 3D case.

\begin{figure}[t]
\begin{center}
\includegraphics[width=0.9\columnwidth,clip]{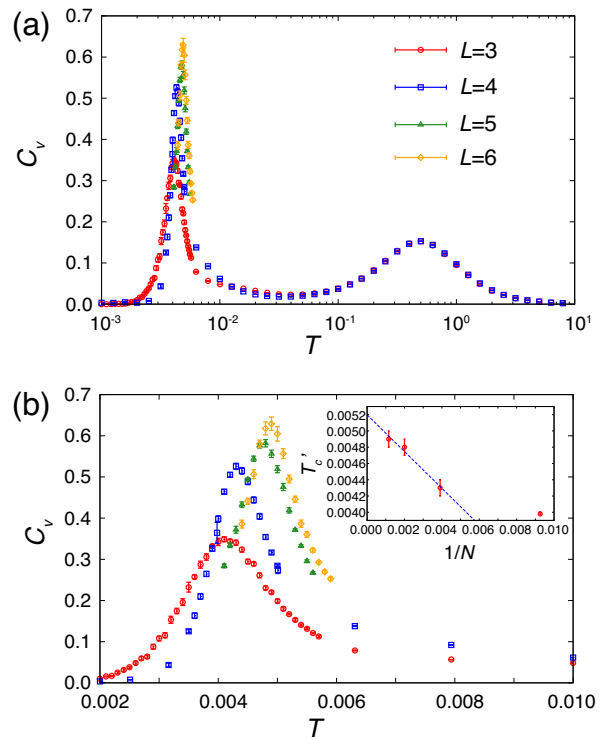}
\caption{(color online).
(a) Temperature dependence of the specific heat in the isotropic case with $J_x=J_y=J_z=1/3$ ($\alpha=1$).
(b) The enlarged view in the vicinity of the low-temperature peak. 
The calculations were performed for the systems on the hyperhoneycomb lattice with $N=4\times L^3$ spins up to $L=6$.
The inset in (b) shows the peak temperature $T_c'$ of the specific heat as a function of the inverse of the system size $N$. The dotted line represents the linear fit for three largest $N$.
}
\label{fig2}
\end{center}
\end{figure}

We investigate the thermodynamic properties of the 3D Kitaev model by adopting a MC simulation. 
Since the model given in Eq.~(\ref{eq:5}) is defined on the bipartite lattices, the conventional quantum MC method on the basis of the Suzuki-Trotter decomposition can be applied at first glance. However, due to the bond-dependent interactions in the Kitaev model, the method suffers from the negative sign problem. 
To avoid the problem, we use an alternative MC method as described below.
By applying the Jordan-Wigner transformation~\cite{Chen2007,Feng2007,Chen2008} and rewriting the resulting spinless fermions by Majorana fermions, the Hamiltonian is written in the form
\begin{align}
 {\cal H}={\rm i}J_x\sum_{x\, \textrm{bonds}}c_w c_b-{\rm i}J_y\sum_{y\, \textrm{bonds}}c_b c_w-{\rm i}J_z\sum_{z\, \textrm{bonds}} \eta_r c_b c_w,
 \label{eq:1}
\end{align}
where $c$ and $\bar{c}$ are the Majorana fermion operators, and $\eta_r={\rm i} \bar{c}_b \bar{c}_w$ are $Z_2$ variables defined on each $z$ bond ($r$ is the bond index), as the eigenvalues are $\pm 1$~\cite{Feng2007}.
As the hyperhoneycomb lattice is bipartite, we term black ($b$) and white ($w$) sites so that, on each $x$ bond, the smaller-(larger-)$i$ site corresponds to the white (black) site, where the numbering for sites are done along chains consisting of $x$ and $y$ bonds, as shown in Fig.~\ref{fig1}(b).
The Hamiltonian in Eq.~(\ref{eq:1}) is a free Majorana fermion system coupled with the $Z_2$ degree of freedom, $\{\eta_r\}$, on each $z$ bond. 
Formally, the model is similar to the double-exchange model with Ising localized spins.
This allows us to apply the MC algorithms developed for the double-exchange models. 
Here, we adopt the conventional algorithm in which the MC weight for a given configuration of $\{\eta_r\}$ is obtained by the exact diagonalization of the Majorana fermions~\cite{Yunoki1998}.
We impose the open boundary conditions for the $\bm{a}$ and $\bm{b}$ directions and the periodic boundary condition for $\bm{c}$ direction to avoid a subtle boundary problem intrinsic to the Jordan-Wigner transformation [see Fig.~\ref{fig1}(b)]. The cluster size $N=4L^3$ in which the calculations are performed is taken up to $L=6$. The details of calculation methods are given in the Supplemental Material~\cite{SM}.

Figure~\ref{fig2}(a) shows the $T$ dependence of the specific heat $C_v$ for the isotropic case with $J_x=J_y=J_z=1/3$.
There are two peaks in $C_v$. The high-$T$ peak at $T \sim 0.6$ does not show the size dependence. On the other hand, the low-$T$ peak located at $T\sim 0.004$ grows with increasing the system size as shown in Fig.~\ref{fig2}(b). This is a signature of phase transition between the low-$T$ QSL phase and the high-$T$ paramagnetic state, as firmly supported by the perturbation arguments below.
The size extrapolation of the peak temperature $T_c'$ gives the estimate of the critical temperature in the thermodynamic limit as $T_c=0.00519(9)$ [see the inset of Fig.~\ref{fig2}(b)]~\cite{fn}. 
In contrast, the 2D Kitaev model does not show such growing peak in the specific heat, indicating the absence of the finite-$T$ phase transition~\cite{SM}.

\begin{figure}[t]
\begin{center}
\includegraphics[width=0.8\columnwidth,clip]{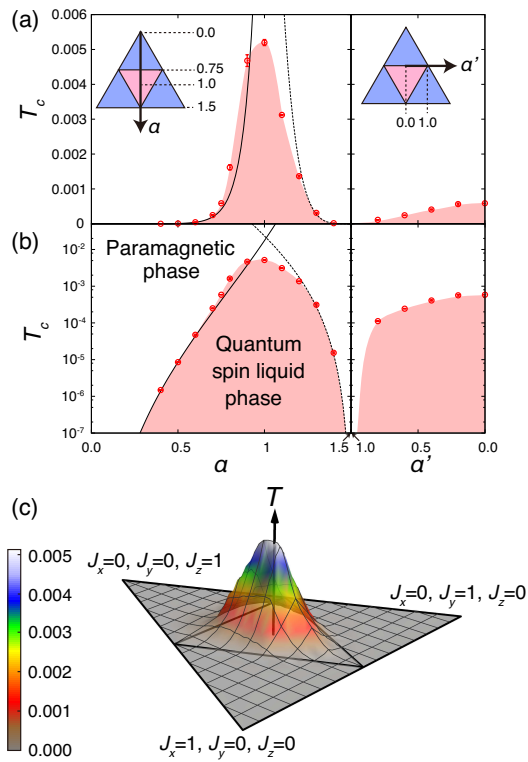}
\caption{(color online).
Finite-temperature phase diagram of the 3D Kitaev model. (A) Cut of the phase diagram along the $\alpha$ and $\alpha'$ axes shown in the insets.
Log-scale plot for (a) is shown in (b). The solid (dashed) line is the $\alpha$ dependence of $T_c$ obtained by the perturbation expansion in terms of $J/J_z$ ($J_z/J$), where $J=J_x=J_y$. 
(c) 3D plot of the phase diagram in the whole parameter space.
The base triangle represents the ground state phase diagram shown in Fig.~\ref{fig1}(c). 
}
\label{fig3}
\end{center}
\end{figure}

By performing the simulation for various sets of $J_x$, $J_y$, and $J_z$, we obtain the finite-$T$ phase diagram of the 3D Kitaev model. 
The results are summarized in Fig.~\ref{fig3}.
Figure~\ref{fig3}(a) shows $T_c$ as a function of the anisotropy parameters $\alpha$ and $\alpha'$ shown in the inset [Figure~\ref{fig3}(b) is the log plot of the same data].
The critical temperature $T_c$ takes the maximum value at $\alpha \simeq 1$ corresponding to the isotropic case, and decreases to zero as $\alpha \to 0$ and $\alpha \to 3/2$. 
The limit of $\alpha \to 0$ corresponds to $J_z \to 1$ with $J_x=J_y=J \to 0$.
This limit was discussed by MC simulation for the effective model obtained by the perturbation theory in terms of $J/J_z$ by the authors and their collaborators~\cite{Nasu2014}. 
A finite-$T$ transition was found at $T_c=\tilde{T}_c\times 7 J^6/(256J_z^5)$ with $\tilde{T}_c=1.925(1)$.
This asymptotic form of $T_c$ is plotted by the solid lines in Figs.~\ref{fig3}(a) and \ref{fig3}(b).
It shows fairly good agreement with the present MC results in the small $\alpha$ region, which strongly supports that $T_c$ estimated from the anomaly in $C_v$ is indeed the critical temperature between the low-$T$ QSL and high-$T$ paramagnet. 
Meanwhile, in the limit of $\alpha \to 3/2$, by using the perturbation expansion in terms of $J_z/J$, we find that $T_c$ is scaled by $J_z^4/J^3$~\cite{SM}. 
The dashed lines in Figs.~\ref{fig3}(a) and \ref{fig3}(b) represent the fitting of MC data by this asymptotic scaling.
It also well explains the MC data, supporting the phase transition at $T_c$.

Figure~\ref{fig3}(c) summarizes the MC estimates of $T_c$ in the 3D plot. 
In the entire parameter space, the low-$T$ QSL is separated from the high-$T$ paramagnet by the thermodynamic singularity at $T_c$. 
There is no adiabatic connection between the two states, and the transition always appears to be continuous within the present calculations. 
These are in sharp contrast to the situation in conventional fluids where liquid and gas are adiabatically connected with each other beyond the critical end point in the phase boundary of the discontinuous transition.
Thus, thermodynamics of the QSLs is not understood by the conventional theory for liquids.

Interestingly, the value of $T_c$ becomes maximum at $\alpha \simeq 1$: the QSL phase is most stable against thermal fluctuations in the isotropic case. The bond-dependent interactions in the Kitaev model compete with each other; it is not possible to optimize the exchange energy on the $x$, $y$, and $z$ bonds simultaneously. 
The frustration becomes strongest at $\alpha=1$.
Hence, interestingly, our MC results in Fig.~\ref{fig3}(c) show that the frustration tends to stabilize the QSL against thermal fluctuations. 
This frustration effect is opposite to that on conventional magnetically ordered states where frustration suppresses the critical temperatures.

\begin{figure}[t]
\begin{center}
\includegraphics[width=\columnwidth,clip]{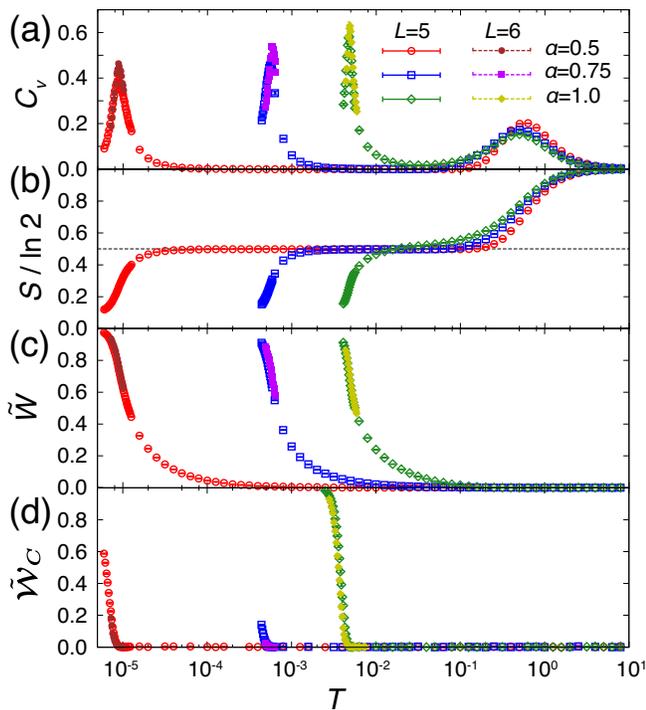}
\caption{(color online).
Temperature dependences of (a) the specific heat, (b) entropy, 
(c) $Z_2$ variables $W_p$ per ten-site plaquette, $\tilde{W}$, and (d) the Wilson loop $\tilde{{\cal W}}_C$.
}
\label{fig4}
\end{center}
\end{figure}

In the vicinity of $\alpha=1$, the ground state is the gapless QSL. By decreasing $\alpha$, the ground state changes into the gapped QSL at the quantum critical point at $\alpha=3/4$, as shown in Fig.~\ref{fig1}(c).
However, $T_c$ changes smoothly around $\alpha=3/4$, as shown in Fig.~\ref{fig3}. 
Also we find no singularity in the $T$ dependence of $C_v$ around $\alpha=3/4$ within the present precision, except for $T_c$ [e.g., see Fig.~\ref{fig4}(a)]. 
In the low-$T$ limit, however, there should be some anomaly in $C_v$, reflecting the change of low-energy excitations. 
The results suggest that such anomaly will happen to be seen at much lower $T$ than $10^{-4}$.

Now let us discuss the reason why the specific heat $C_v$ exhibits two peaks.
We show the $T$ dependence of the entropy per site, $S$, in Fig.~\ref{fig4}(b), obtained by the numerical integration of $C_v$ divided by $T$.
By decreasing $T$, the entropy decreases from $\ln 2$ corresponding to the high-$T$ peak in $C_v$ and approaches $\frac12 \ln 2$.
In the $T$ region between the two peaks in $C_v$, the entropy stays at $\simeq \frac12 \ln 2$.
As further decreasing $T$, the entropy rapidly deceases again corresponding to the low-$T$ peak in $C_v$, and approaches zero toward $T=0$.
The successive entropy release is ascribed to a separation of the energy scales for the Majorana fermions and the $Z_2$ variables $\eta_r$. 
Namely, while decreasing $T$, the entropy of $\frac12 \ln 2$ associated with Majorana fermions is first gradually released at $T \sim 0.1-1$, corresponding to their kinetic energy scale $\sim J_x+J_y+J_z =1$. 
Subsequently, the remaining entropy of $\frac12 \ln 2$, associated with the $Z_2$ variables, is released at the phase transition.
This lower energy scale is set by the effective interactions between the $Z_2$ variables mediated by Majorana fermions, which depend on the anisotropy of the system.
We confirm this picture by calculating $\tilde{W}$ defined as the thermal average of the density of the $Z_2$ variables $W_p=\pm 1$ defined for each ten-site loop [see Fig.~\ref{fig1}(b)], which is computed by the product of $\eta_r$~\cite{SM}.
Figure~\ref{fig4}(c) shows the $T$ dependence of $\tilde{W}$. This quantity rapidly increases at the lower-$T$ peak in $C_v$ as $T$ decreases. 
Therefore, the entropy of $\frac12 \ln 2$ is released according to the coherent growth of $W_p$ at $T_c$.

However, it is worth noting that the phase transition at $T_c$ is not caused by the symmetry breaking in terms of the local variables $W_p$. 
Instead, the phase transition will be understood by the topological nature of excited states as follows.
The excited states are generated by flipping $W_p$ from the ground state where all $W_p=+1$. 
The flipped $W_p=-1$ form loops because of the local constraints originating from the fundamental spin-1/2 algebra~\cite{Mandal2009}.
The excitation energy of loops and their configurational entropy compete with each other, which may lead to the phase transition at a finite $T$, as is discussed by Peierls for the 2D Ising model~\cite{Peierls1936}. 
This picture was indeed confirmed in the limit of $J_z\gg J_x,J_y$, through the winding number defined for $W_p$~\cite{Nasu2014}.
In the present case, however, the winding number cannot be defined, as the calculations are done under the open boundary conditions in the $a$ and $b$ directions.
Instead, we calculate the thermal average of the Wilson loop along the edge of the $ab$ plane, $\tilde{{\cal W}}_C$, which serves as an alternative parameter to the winding number~\cite{SM}. As shown in Fig.~\ref{fig4}(d), $\tilde{{\cal W}}_C$ behaves like an order parameter: it becomes nonzero below $T_c$~\cite{fn}.
The situation is in sharp contrast to the 2D Kitaev model, where the excitation with $W_p=-1$ is allowed independently without local constraints, and consequently, the QSL is adiabatically connected to the high-$T$ paramagnet.

Our results on the topological transition suggest a new paradigm of critical phenomena beyond the Ginzburg-Landau-Wilson (GLW) theory.
Due to the lack of local order parameter, the description based on the GLW theory is no longer applicable to the ``vaporization'' of QSLs. 
Such nontrivial finite-$T$ phase transitions have been studied by the mean-field approximations for 3D $Z_2$ QSLs on the basis of the $Z_2$ gauge theory~\cite{Senthil2000,Wen2002}. 
To understand the critical properties, however, it is necessary to take into account fluctuations of a topological structure in the excitations beyond the mean-field approach. 
The current study presents the first unbiased results on topological transitions, which may give birth to a new concept of critical phenomena beyond the conventional GLW theory.

It will also be interesting to consider the ``solidification'' of QSLs. Indeed, the solid phase (magnetically ordered phase) is accessible in the context of the present 3D Kitaev model, by considering additional interactions which favor a magnetic order, such as the Heisenberg exchange interaction~\cite{Lee2014,Kimchi1309}.
The detailed study of the magnetic three states of matter, liquid, gas, and solid, will provide a new insight in the research area of magnetism.

The present results give a counterexample to the conventional ``myth'' on QSLs: 
the absence of phase transition is a requirement for QSL.
This myth has long haunted the experimental identification of QSLs.
Our results, however, indicate that a phase transition does not always signal symmetry breaking by a magnetic long-range order.
This will urge reconsideration of the experimental detection of QSLs; even if the system exhibits a phase transition, it should not be excluded from the candidates for QSLs, as long as a clear indication of magnetic ordering is not established.

\begin{acknowledgments}
We thank L. Balents, M. Imada, and O. Tchernyshyov for fruitful discussions.
J.N. is supported by the Japan Society for the Promotion of Science through a research fellowship for young scientists.
This work is supported by Grant-in-Aid for Scientific Research, the Strategic Programs for Innovative Research (SPIRE), MEXT, and the Computational Materials Science Initiative (CMSI), Japan.
Parts of the numerical calculations are performed in the supercomputing systems in ISSP, the University of Tokyo.
\end{acknowledgments}



\clearpage

\section*{\Large 
Supplemental Material for the article\\
``Vaporization of Kitaev spin liquids''
}

\section*{\large Calculation Method}
\label{app:sec:method}

In this section, we present the method for calculating the thermodynamic quantities in the Kitaev model. 
The method is commonly used for the 2D honeycomb and 3D hyperhoneycomb lattices.
The Kitaev model is given by 
\begin{align}
{\cal H}=-J_x\sum_{\langle ij\rangle_x}\sigma_i^x\sigma_j^x-J_y\sum_{\langle ij\rangle_y}\sigma_i^y\sigma_j^y-J_z\sum_{\langle ij\rangle_z}\sigma_i^z\sigma_j^z,\label{supp_eq:5}
\end{align}
where $\sigma_i^x$, $\sigma_i^y$, and $\sigma_i^z$ are Pauli matrices describing a spin-1/2 state at a site $i$;
$J_x$, $J_y$, and $J_z$ are exchange constants defined on three different types of the nearest neighbor bonds, $x$, $y$, and $z$ bonds, respectively.
Since this model is defined on the bipartite lattices, the conventional quantum Monte Carlo (MC) method on the basis of the Suzuki-Trotter decomposition can be applied at first glance. However, due to the bond-dependent interactions in the Kitaev model, the method suffers from the negative sign problem. 
To avoid the problem, we use an alternative MC method as described below.

First, we regard the honeycomb and hyperhoneycomb lattices as assemblies of 1D chains composed of $x$ and $y$ bonds, and associate each site $i$ with a pair of integers, $(m,n)$. Here, $m$ allocates a chain and $n$ is the site index within the $m$-th chain.
By applying the Jordan-Wigner transformation~\cite{supp_Chen2007,supp_Feng2007,supp_Chen2008}, the spin operators are written by spinless fermion operators ($a_i$, $a_i^\dagger$) as
\begin{align}
 S_{m,n}^+&=(S_{m,n}^-)^\dagger=\frac{1}{2}(\sigma_{m,n}^x+i \sigma_{m,n}^y)\nonumber\\
&=\prod_{n'=1}^{n-1}(1-2n_{m,n'}) a_{m,n}^\dagger,\label{supp_eq:3}\\
\sigma_{m,n}^z&=2n_{m,n}-1,\label{supp_eq:4}
\end{align}
where $n_i$ is the number operator defined by $n_i=a_{i}^\dagger a_{i}$.
Then, the interactions in Eq.~(\ref{supp_eq:5}) are written as
\begin{align}
\label{supp_eq:xx}
& \sigma_{m,n}^x\sigma_{m,n+1}^x=-(a_{m,n}-a_{m,n}^\dagger)(a_{m,n+1}+a_{m,n+1}^\dagger),\\
\label{supp_eq:yy}
& \sigma_{m,n}^y\sigma_{m,n+1}^y=(a_{m,n}+a_{m,n}^\dagger)(a_{m,n+1}-a_{m,n+1}^\dagger), \\
\label{supp_eq:zz}
& \sigma_{m,n}^z\sigma_{m',n'}^z=(2n_{m,n}-1)(2n_{m',n'}-1).
\end{align}
As both the honeycomb and hyperhoneycomb lattices are bipartite, we term black ($b$) and white ($w$) sites so that, on each $x$ bond, the smaller-(larger-)$n$ site corresponds to the white (black) site, as shown in Figs.~\ref{fig_hyperhoneycomb} and~\ref{fig_honeycomb}.
By using Eqs.~(\ref{supp_eq:xx})-(\ref{supp_eq:zz}), the Hamiltonian in Eq.~(\ref{supp_eq:5}) is rewritten as
\begin{align}
{\cal H}=J_x\sum_{x\, \textrm{bonds}}(a_w-a_w^\dagger)(a_b+a_b^\dagger)\nonumber\\
-J_y\sum_{y\, \textrm{bonds}}(a_b+a_b^\dagger)(a_w-a_w^\dagger)\nonumber\\
-J_z\sum_{z\, \textrm{bonds}}(2n_b-1)(2n_w-1).
\end{align}
Next, we introduce Majorana fermion operators $c$ and $\bar{c}$ from the spinless fermion operators as
\begin{align}
& c_w=(a_w-a_w^\dagger)/{\rm i},\ \ \ \bar{c}_w=a_w+a_w^\dagger,\\
& c_b=a_b+a_b^\dagger,\ \ \ \bar{c}_b=(a_b-a_b^\dagger)/{\rm i}.
\end{align}
By using the Majorana fermion operators, the Hamiltonian is written in the form
\begin{align}
 {\cal H}={\rm i}J_x\sum_{x\, \textrm{bonds}}c_w c_b-{\rm i}J_y\sum_{y\, \textrm{bonds}}c_b c_w-{\rm i}J_z\sum_{z\, \textrm{bonds}} \eta_r c_b c_w,
 \label{supp_eq:1}
\end{align}
where $\eta_r={\rm i} \bar{c}_b \bar{c}_w$ are $Z_2$ variables defined on each $z$ bond ($r$ is the bond index), as the eigenvalues are $\pm 1$~\cite{supp_Feng2007}.
Here, we consider that 1D chains composed of $x$ and $y$ bonds are open strings, by imposing open boundary conditions along the chain, in order to avoid a subtle boundary problem intrinsic to the Jordan-Wigner transformation.
Under periodic boundary conditions, a complicated nonlocal term depending on the parity of the total fermion number will appear at the boundary from Eq.~(\ref{supp_eq:3}). 

The Hamiltonian in Eq.~(\ref{supp_eq:1}) is a free Majorana fermion system coupled with the $Z_2$ degree of freedom, $\{\eta_r\}$, on each $z$ bond. 
Formally, the model is similar to the double-exchange model with Ising localized spins; in the usual double-exchange models, localized spins couple with itinerant electron spins via the on-site exchange coupling, but in the present case, the Ising spins couple with the hopping of fermions along the $z$ bonds. 
The formal equivalence allows us to apply the MC algorithms developed for the double-exchange models. 
Here, we adopt the conventional algorithm in which the MC weight for a given configuration of $\{\eta_r\}$ is obtained by the exact diagonalization of the Majorana fermions~\cite{supp_Yunoki1998}, as described below.

\begin{figure}[t]
\begin{center}
\includegraphics[width=0.8\columnwidth,clip]{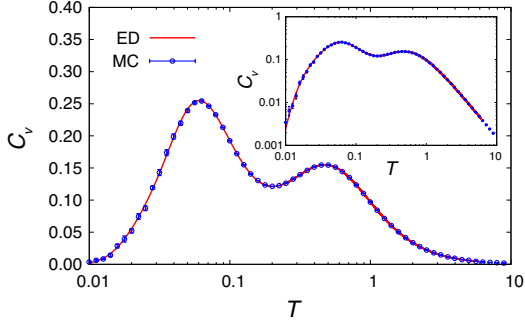}
\caption{
Benchmark of the quantum Monte Carlo method.
The MC result for the Majorana-fermion Hamiltonian in Eq.~(\ref{supp_eq:1}) and the exact diagonalization (ED) result for the original quantum-spin Hamiltonian in Eq.~(\ref{supp_eq:5}) are presented for comparison.
The inset shows the log-log plot of the same data.
These are obtained for the 2D Kitaev model on the honeycomb lattice cluster with $2\times 2^2$ sites by adopting the type-II boundary condition in Fig.~\ref{fig_honeycomb}(a).
The parameters are chosen to be $J_x=J_y=J_z=1/3$.
}
\label{2DCv_hikaku}
\end{center}
\end{figure}

The partition function of the system described by the Hamiltonian in Eq.~(\ref{supp_eq:1}) is given by 
\begin{align}
 Z={\rm Tr}_{\{\eta_r\}}{\rm Tr}_{\{c_i\}}e^{-\beta {\cal H}}={\rm Tr}_{\{\eta_r\}}e^{-\beta F_{f}(\{\eta_r\})},\label{supp_eq:2}
\end{align}
where $\beta$ is the inverse temperature $\beta=1/T$ (we set the Boltzmann constant $k_{\rm B}=1$).
 $F_{f}(\{\eta_r\})$ is the free energy of the Majorana fermion system for a given configuration of $\{\eta_r\}$;
\begin{align}
\label{supp_eq:F_f}
 F_{f}(\{\eta_r\})=-T \ln {\rm Tr}_{\{c_i\}}e^{-\beta {\cal H}(\{\eta_r\})}.
\end{align}
For a given $\{\eta_r\}$, the quadratic Hamiltonian ${\cal H}(\{\eta_r\})$ is easily diagonalized to give
\begin{align}
{\cal H}(\{\eta_r\})=\sum_\lambda^{N/2}\varepsilon_\lambda(\{\eta_r\})\left(f_\lambda^\dagger f_\lambda -\frac{1}{2}\right),
\end{align}
where $f_\lambda$ ($f_\lambda^\dagger$) is the annihilation (creation) operator of a spinless fermion [see also Eqs.~(\ref{supp_eq:H_1D_diag}) and (\ref{supp_eq:def_f})]. 
It is worthy noting that there are $N/2$ one-body states in the Majorana fermion for the $N$-site system.
Then, the free energy is calculated as
\begin{align}
 F_{f}(\{\eta_r\})=-T\sum_\lambda \ln [2\cosh (\beta\varepsilon_\lambda /2)].
\end{align}
We perform the Markov-chain MC simulation for the classical local variables $\eta_r=\pm 1$ so as to reproduce the Boltzmann distribution of $e^{-\beta F_{f}(\{\eta_r\})}$. The energy $E$ and the specific heat $C_v$ at the temperature ($T$) are calculated as
\begin{align}
 E&=\means{E_f}_{\rm MC},\\
C_v&=\frac{\partial E}{\partial T}=\frac{1}{T^2}
\left(
\means{E_f^2}_{\rm MC}-\means{E_f}^2_{\rm MC}-\mean{\frac{\partial E_f}{\partial \beta}}_{\rm MC}
\right),
\end{align}
respectively. Here, $E_f$ is the internal energy of the Majorana fermion system given by
\begin{align}
 E_f(\{\eta_r\})=-\sum_\lambda\frac{\varepsilon_\lambda}{2}\tanh\frac{\beta\varepsilon_\lambda}{2},
\end{align}
and $\means{\cdots}_{\rm MC}$ represents the thermal average calculated by the MC simulation.
The validity of the MC simulation for the Majorana-fermion Hamiltonian in Eq.~(\ref{supp_eq:1}) was confirmed by comparing the results with those for the original quantum-spin Hamiltonian in Eq.~(\ref{supp_eq:5}). 
We performed the comparison in the small size clusters, for which the results for the original Hamiltonian can be obtained by the exact diagonalization. 
(MC simulation for the original Hamiltonian is suffered from the negative sign problem, as mentioned above.) 
Figure~\ref{2DCv_hikaku} shows the comparison of the specific heat obtained by the two methods for the $L=2$ cluster ($2\times 2^2 = 8$ sites) in the 2D Kitaev model. 
The MC results well reproduce the exact diagonalization results within the statistical errors in the entire $T$ range. This indicates that the present MC simulation for the Majorana fermion system with the Ising degree of freedom in Eq.~(\ref{supp_eq:1}) gives numerically-exact results for the thermodynamics of the Kitaev model given in Eq.~(\ref{supp_eq:5}).

\section*{\large Details of Monte Carlo simulation}

\subsection*{Replica  exchange MC simulation}
\label{app:sec:replica}

The replica exchange MC technique is an efficient way to avoid the slowing down and freezing of the MC sampling at low $T$~\cite{supp_Swendsen1986,supp_Hukushima1996}.
In this technique, we prepare several replicas with different temperatures. 
In each replica, we perform a single-flip MC simulation to which the Metropolis algorithm is applied.
In addition, we swap two replicas at fixed intervals of the single-flip MC samplings.
As our system includes fermions, we need a modification for the swap procedure to the standard one for localized spin systems. 
We describe the modification below.

The standard replica exchange for localized spin systems~\cite{supp_Swendsen1986,supp_Hukushima1996} is performed so that the exchange probability $p$ between a replica with $\{\eta_r\}_i$ at the temperature $T_i=1/\beta_i$ and another replica with $\{\eta_r\}_j$ at the temperature $T_j=1/\beta_j$ is given by
\begin{align}
 p={\rm min}(1,f),\label{supp_eq:11}
\end{align}
with 
\begin{align}
  f&=\frac{\exp\left[-\beta_i E(\{\eta_r\}_j) - \beta_j E(\{\eta_r\}_i)\right]}{\exp\left[-\beta_i E(\{\eta_r\}_i) - \beta_j E(\{\eta_r\}_j)\right]}\nonumber\\
&=\exp\left[(\beta_i-\beta_j)(E(\{\eta_r\}_i)-E(\{\eta_r\}_j))\right].
\label{supp_eq:13}
\end{align}
Here, $E(\{\eta_r\}_i)$ is the energy in the replica with $\{\eta_r\}_i$, which is not dependent on $T$. 
The probability in Eq.~(\ref{supp_eq:11}) satisfies the detailed balance so as to reproduce the Boltzmann distribution for the whole system including all the replicas.

In contrast, in the present model in Eq.~(\ref{supp_eq:1}) including fermions, the Boltzmann weight is given by the fermion free energy in Eq.~(\ref{supp_eq:F_f}),
 which depends on $T$. Hence, we need to modify $f$ in Eq.~(\ref{supp_eq:13}) as
\begin{align}
f=\exp[&-\beta_i F_f(\beta_i,\{\eta_r\}_j) - \beta_j F_f(\beta_j,\{\eta_r\}_i)\nonumber\\
 &+\beta_i F_f(\beta_i,\{\eta_r\}_i) + \beta_j F_f(\beta_j,\{\eta_r\}_j)].\label{supp_eq:12}
\end{align}
Thus, for calculating $f$, we need additional calculations to evaluate $F_f(\beta_j,\{\eta_r\}_i)$ and $F_f(\beta_j,\{\eta_r\}_i)$ (note the subsrtipts $i$ and $j$).

The replica exchange is very effective for avoiding the slowing down at low $T$ in the present calculations; in fact, the acceptance ratio of the replica exchange process becomes rather higher than that of the single flip process at low $T$, as mentioned in the next section.
Moreover, the calculation with the replica exchange is suitable for a parallel computation.

\subsection*{Conditions of MC simulations}
\label{app:sec:conditions}

\begin{figure}[t]
\begin{center}
\includegraphics[width=0.6\columnwidth,clip]{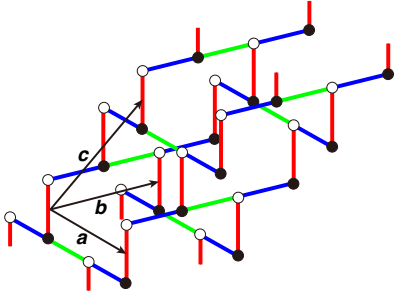}
\caption{
Hyperhoneycomb lattice on a $4\times L^3$ cluster with $L=2$.
Open boundary conditions are imposed in the $\bm{a}$ and $\bm{b}$ directions and a periodic boundary condition is imposed in the $\bm{c}$ direction.
}
\label{fig_hyperhoneycomb}
\end{center}
\end{figure}

All the simulations for the 3D model in the main text were done in the following conditions.
The calculations were done for the $N=4L^3$ site cluster with open boundary conditions for the $\bm{a}$ and $\bm{b}$ directions and a periodic boundary condition for the $\bm{c}$ direction, as shown in Fig.~\ref{fig_hyperhoneycomb}.
Here, $N$ is the number of lattice sites.
Typically, we prepared $N_r=16$ replicas, and performed 40,000 (16,000) MC steps for measurements after 10,000 (1,000) MC steps for thermalization for the $L=3,4,5$ clusters ($L=6$ cluster).
Here, one MC step includes $N$-times trials of single flips of $\eta_r$ in each replica and $N_r$-times exchanges of a pair of replicas with neighbouring temperatures.
Typical acceptance ratios for a single flip and a replica exchange are about $10\%$ and $60\%$, respectively, in the vicinity of the low-$T$ peak in the specific heat.

\subsection*{Calculation of the local conserved quantity}
\label{app:sec:calc_local}

We computed the local conserved quantity $W_p$ defined on each ten-site loop $p$ on the hyperhoneycomb lattice. This quantity is given by 
\begin{align}
 W_p=\prod_{i\in p}\sigma_{i}^{l_{i}}.
 \label{supp_eq:10}
\end{align}
Here, $l_i=x$, $y$, or $z$ is one of the three bonds at site $i$ that is not included in the loop $p$.
By substituting the Jordan-Wigner representations, $W_p$ is rewritten by a product of the $Z_2$ variables $\eta_r$ included in the loop $p$, as
\begin{align}
 W_p=\prod_{r\in p}\eta_r.
\end{align}
Since there are four $z$ bonds on a ten-site loop, $W_p$ is given by the product of four $\eta_r$.
We computed the quantity $\tilde{W}=\sum_p\langle W_p\rangle/N_p$ in the main article by taking the thermal average of $W_p$ in this form by using the MC calculation.
Here, $N_p$ is the number of ten-site loops in the hyperhoneycomb cluster.

\subsection*{Calculation of Wilson loop}
\label{app:sec:calc_wilson}

\begin{figure}[t]
\begin{center}
\includegraphics[width=0.87\columnwidth,clip]{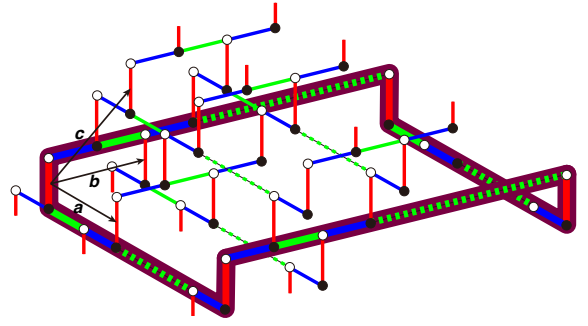}
\caption{
A closed contour $C$ on which the Wilson loop ${\cal W}_C$ in Eq.~(\ref{supp_eq:15}) is defined (thick line).
}
\label{edge}
\end{center}
\end{figure}

The conserved quantity $W_p$ defined on a ten-site loop in Eq.~(\ref{supp_eq:10}) gives a shortest Wilson loop. In more general form, the Wilson loop is defined along a closed contour ${\cal C}$ as~\cite{supp_Kitaev06,supp_Mandal2009}
\begin{align}
 {\cal W}_{{\cal C}}=\prod_{i\in {\cal C}}\sigma_{i}^{l_{i}}.
 \label{supp_eq:6}
\end{align}
Paying attention to the open boundary conditions along the $\bm{a}$ and $\bm{b}$ directions imposed in the present study, we here calculate the Wilson loop ${\cal W}_C$ on the contour $C$ along the edge of an $ab$ plane as shown in Fig.~\ref{edge}.
In this case, ${\cal W}_C$ is rewritten by using $W_p$ and $\eta_r$ into
\begin{align}
 {\cal W}_C=-\prod_{p\in S_C}W_p = -\prod_{r\in C}\eta_r\equiv -{\cal W}'_C,
\label{supp_eq:15}
\end{align}
where $S_C$ is the $ab$ plane surrounded by the contour $C$. Note that the negative sign appears due to the relation $\sigma^x \sigma^y=-\sigma^y \sigma^x= i\sigma^z$ and their cyclic permutations. Here, we choose the sign of ${\cal W}'_C$ so as to satisfy ${\cal W}'_C=+1$ in the ground state.
In the MC calculations, we compute the thermal average of ${\cal W}'_C$, together with taking the average over all the $ab$ slices, in the form
\begin{align}
 \tilde{{\cal W}}_C=\frac{1}{L}\sum_{i=1}^{L}\langle {\cal W}'_{C_i}\rangle,
 \label{supp_eq:W_C_MC}
\end{align}
where $C_i$ is the the contour along the edge of $i$-th $ab$ plane.

$\tilde{{\cal W}}_C$ in Eq.~(\ref{supp_eq:W_C_MC}) is a candidate of the ``order parameter'' for the phase transition without apparent symmetry breaking studied in the main text, as explained in the following.
The Wilson loop ${\cal W}_C$ along the edge of an $ab$ plane is related to the winding number of the loops composed of flipped $W_p$ introduced in Ref.~\cite{supp_Nasu2014}. As discussed in Ref.~\cite{supp_Nasu2014}, the winding number is successfully used to characterize the phase transition between QSL and paramagnet in the case of $J_z\gg J_x,J_y$ in the MC calculations under the periodic boundary conditions: it is nonzero in the high-$T$ paramagnetic phase but continuously vanishes in entering into the low-$T$ QSL phase. This behavior is expected from the fact that only short loops are excited below $T_c$ and the loops extending from one edge of the system to the other, which contribute the winding number, are not excited below $T_c$. 
The Wilson loop ${\cal W}_C$ along the edge of the $ab$ plane represents the parity of the number of $W_p=-1$ on the $ab$ plane from Eq.~(\ref{supp_eq:15}), namely, the number of the loops of $W_p=-1$ intersecting the $ab$ plane.
When a short loop intersects the $ab$ plane, it will lead to ${\cal W}'_C=+1$. On the other hand, when extended loops are thermally excited, they will contribute to both ${\cal W}'_C=\pm 1$ with equal weights. Therefore, we expect that $\tilde{{\cal W}}_C$ takes a nonzero value below $T_c$ and vanishes above $T_c$.

\section*{\large Thermodynamic properties in 2D Kitaev model}

Here, we present the MC results for the 2D Kitaev model on a honeycomb lattice. 
Focusing on the isotropic case with $J_z=J_y=J_z=1/3$, we show that the MC data indicate no phase transition at finite $T$.

\subsection*{Boundary conditions}
\label{app:sec:2D_bc}

\begin{figure}[t]
\begin{center}
\includegraphics[width=0.87\columnwidth,clip]{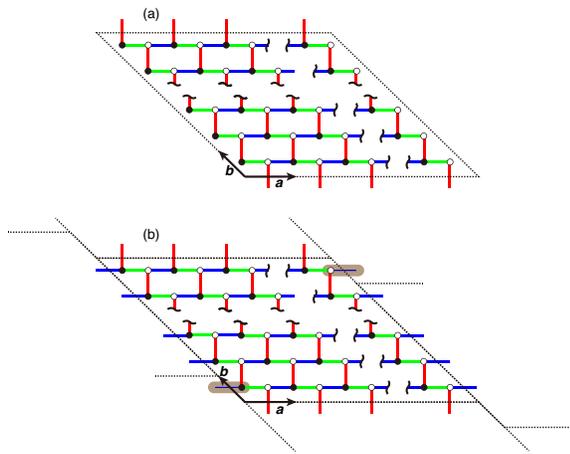}
\caption{
Boundary conditions for a finite-size cluster of the honeycomb lattice.
(a)~Type-I boundary condition: open boundary condition in the $\bm{a}$ direction and periodic boundary condition in the $\bm{b}$ direction.
(b)~Type-II boundary condition.  We neglect the interactions on the shaded bond connecting the lower-left and the upper-right sites.
The system is regarded as a single open chain consisting of $x$ and $y$ bonds.
}
\label{fig_honeycomb}
\end{center}
\end{figure}

In the present calculation, we assume two different types of boundary conditions. One is the type-I boundary condition as shown in Fig.~\ref{fig_honeycomb}(a).
This is a 2D analogue of that for the 3D hyperhoneycomb lattice in Fig.~\ref{fig_hyperhoneycomb}: 
we impose the open boundary condition along the $\bm{a}$ direction and the periodic boundary condition along the $\bm{b}$ direction. 
In this case, the system is regarded as a bundle of many open chains.
We also consider the other type of the boundary condition, as shown in Fig.~\ref{fig_honeycomb}(b). 
In this lattice geometry, the system is considered as a single open chain, terminated at the lower-left and the upper-right sites. 
Namely, we impose a shifted periodic boundary condition in the $\bm{a}$ direction and omit the interaction on the $x$ bonds connecting the lower-left and the upper-right of the cluster 
[the shaded bond in Fig.~\ref{fig_honeycomb}(b)]. 
Along the $\bm{b}$ direction, we impose the periodic boundary condition.
We term this latter case the type-II boundary condition.
Note that the boundary problem intrinsic to the Jordan-Wigner transformation does not show up for both types of the boundary conditions.

\subsection*{Calculation results}
\label{app:sec:2D_results}

\begin{figure}[t]
\begin{center}
\includegraphics[width=0.8\columnwidth,clip]{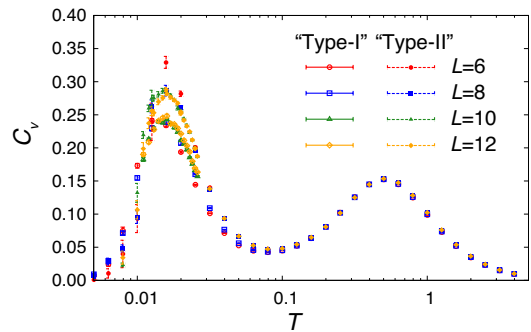}
\caption{
Temperature dependence of the specific heat for the 2D Kitaev model.
``Type-I'' (``Type-II'') represents the results with the type-I (type-II) boundary condition shown in Fig.~\ref{fig_honeycomb}(a)
(Fig.~\ref{fig_honeycomb}(b)).
}
\label{2DCv}
\end{center}
\end{figure}

Figure~\ref{2DCv} shows the $T$ dependence of the specific heat obtained by the MC simulation with two different types of the boundary conditions on finite-size clusters with $2L^2$ sites ($L=6-12$). 
The specific heat exhibits two peaks, as in the 3D case shown in the main article.
However, the low-$T$ peak does not grow as the system size increases, in contrast to the 3D case. 
The two series for different boundary conditions appear to converge to a broad peak with a finite peak height in the thermodynamic limit.
Thus, both peaks are crossover, and there is no singularity in the specific heat in the 2D Kitaev model, in contrast to that in the 3D Kitaev model. The result indicates the absence of the finite-$T$ phase transition in the 2D case.
Note that the absence of phase transition is rigorously shown in the limit of $J_z\gg J_x,J_y$ in the 2D Kitaev model (the toric code limit)~\cite{supp_Castelnovo2007,Nussinov2008}.

\section*{\large Perturbation expansions}

In order to perform perturbation expansions, we divide the Kitaev model into the following two terms:
\begin{align}
 {\cal H}_{xy}&=-J_x\sum_{\langle ij\rangle_x}\sigma_i^x\sigma_j^x-J_y\sum_{\langle ij\rangle_y}\sigma_i^y\sigma_j^y,\\
 {\cal H}_{z}&=-J_z\sum_{\langle ij\rangle_z}\sigma_i^z\sigma_j^z.
\end{align}
We perform the perturbation expansion from two different limits: $J_x, J_y\ll  J_z$ and $J_z\ll J_x, J_y$.

\subsection*{Perturbation expansion for $J_x, J_y\ll  J_z$}

In this section, we briefly review the results of a perturbation expansion where we regard ${\cal H}_{xy}$ as the perturbation term.
This expansion was performed in Ref.~\cite{supp_Mandal2009} and the thermodynamic properties were investigated in Ref.~\cite{supp_Nasu2014}. 
As shown in Ref.~\cite{supp_Mandal2009}, the lowest-order nonzero contribution in the perturbation expansion appears in the sixth order of $J_x$ and $J_y$; an effective Ising-type model was derived with a coupling constant $\propto J^6/J_z^5$, where we take $J=J_x=J_y$. 
Finite-$T$ properties of the effective model was numerically studied in Ref.~\cite{supp_Nasu2014}, and a finite-$T$ phase transition was found at the critical temperature 
\begin{align}
T_c= \frac{7}{256}\frac{J^6}{J_z^5} \times 1.925(1). 
\end{align}
This result is used in the main text.

\subsection*{Perturbation expansion for $J_z\ll J_x, J_y$}
\label{sec:perturb2}

In this section, we perform a perturbation expansion for ${\cal H}={\cal H}_{xy}+{\cal H}_{z}$ by regarding ${\cal H}_{z}$ as the perturbation term.
The perturbation expansion for the free energy $F$ in ${\cal H}$ is given by
\begin{align}
 F=F_0-T\left\{\mean{T_\tau\exp\left[-\int_0^\beta d\tau {\cal H}_{z}(\tau)\right]}_{0c}-1\right\},\label{supp_eq:14}
\end{align}
where $F_0$ is the free energy in the unperturbed Hamiltonian ${\cal H}_{xy}$. 
Here, $\mean{\cdots}_{0c}$ represents the statistical average with ${\cal H}_{xy}$, where only connected diagrams are taken into account. The time-ordering operator is represented as $T_\tau$ for imaginary time $\tau$.
In order to evaluate Eq.~(\ref{supp_eq:14}), we start from calculating Green's functions in the unperturbed Hamiltonian given by [see Eq.~(\ref{supp_eq:1})]
\begin{align}
 {\cal H}_{xy}={\rm i}J_x\sum_{x\, \textrm{bonds}}c_w c_b-{\rm i}J_y\sum_{y\, \textrm{bonds}}c_b c_w.
\end{align}
Since this Hamiltonian consists of independent chains, ${\cal H}_{xy}$ can be given as a set of single-chain Hamiltonian ${\cal H}_{\rm 1D}$ in the form
\begin{align}
 {\cal H}_{\rm 1D}=\sum_{l=1}^{{\cal L}}({\rm i}J_xc_{lw} c_{lb}-{\rm i}J_y c_{lb} c_{l+1w}),
\end{align}
where each chain includes ${\cal L}$ unit cells, and each unit cell includes two sites $w$ and $b$.
We define the Fourier transformation in the 1D chain by
\begin{align}
\label{Fourier1}
 c_{k\gamma}&=\frac{1}{\sqrt{2{\cal L}}} \sum_{l} e^{{\rm i} k r_l} c_{l\gamma},\\
\label{Fourier2}
c_{l\gamma}&=\sqrt{\frac{2}{{\cal L}}} \sum_{k} e^{{\rm i} k r_l} c_{k\gamma},
\end{align}
where $\gamma(=w,b)$ is the sublattice index. The operator $c_{k\gamma}$ behaves as a fermion operator because a relation $\{c_{k\gamma},c_{k'\gamma'}^\dagger\}=\delta_{kk'}\delta_{\gamma\gamma'}$ is satisfied due to the commutation relation $\{c_{l\gamma},c_{l'\gamma'}^\dagger\}=2\delta_{ll'}\delta_{\gamma\gamma'}$ of Majorana fermions. Although $2{\cal L}$ fermions appear to exist in ${\cal H}_{\rm 1D}$ at a first glance, there is an additional relation $c_{-k\gamma}=c_{k\gamma}^\dagger$ which indicates that there are ${\cal L}$ independent fermions in this system.
We here assume the antiperiodic boundary condition to eliminate $k=0$ in the Fourier series in Eqs.~(\ref{Fourier1}) and (\ref{Fourier2}) for avoiding the subtlety arising from the Majorana nature of $k=0$ operator.
By applying a unitary transformation, the 1D Hamiltonian is diagonalized as
\begin{align}
\label{supp_eq:H_1D_diag}
 {\cal H}_{\rm 1D}=\sum_k\tilde{\varepsilon}_k\left(\tilde{f}_k^\dagger \tilde{f}_k-\frac{1}{2}\right),
\end{align}
where $\tilde{\varepsilon}_k$ is given by
\begin{align}
 \tilde{\varepsilon}_k=2\sqrt{J_x^2+J_y^2+2J_x J_y \cos k}.
\end{align}
The fermion operator $\tilde{f}_k$ is given by
\begin{align}
\label{supp_eq:def_f}
 \tilde{f}_k=\frac{1}{\sqrt{2}}c_{kw}+\frac{e^{-{\rm i}\theta_k}}{\sqrt{2}}c_{kb},
\end{align}
where $\theta_k$ is determined so as to be
\begin{align}
 e^{{\rm i}\theta_k}=\frac{2J_y\sin k-2{\rm i}(J_x+J_y\cos k)}{\tilde{\varepsilon}_k}.
\end{align}
Since $e^{-{\rm i}\theta_{-k}}=-e^{{\rm i}\theta_{k}}$, $\tilde{f}_k$ is independent of $\tilde{f}_{-k}^\dagger$.
Then, the operators $\tilde{f}_k$ satisfy the commutation relations: $\{\tilde{f}_{k},\tilde{f}_{k'}^\dagger\}=\delta_{kk'}$ and $\{\tilde{f}_{k},\tilde{f}_{k'}\}=0$.
Green's function for $\tilde{f}_k$ is given as
\begin{align}
 {\cal G}^0_k(i\nu_n)&=-\int_0^\beta d\tau\means{\tilde f_k(\tau)\tilde f_k^\dagger}_0 e^{{\rm i}\nu_n \tau}=\frac{1}{{\rm i}\nu_n-\tilde{\varepsilon}_k},\label{supp_eq:7}\\
 \bar{\cal G}^0_k(i\nu_n)&=-\int_0^\beta d\tau\means{\tilde f_k^\dagger(\tau)\tilde f_k}_0 e^{{\rm i}\nu_n \tau}=\frac{1}{{\rm i}\nu_n+\tilde{\varepsilon}_k},\label{supp_eq:8}
\end{align}
where $\nu_n=(2n+1)\pi T$ is the Matsubara frequency.
The time-dependent operator ${\cal O}(\tau)$ in the interaction representation is given by
\begin{align}
 {\cal O}(\tau)=e^{-\tau{\cal H}_{xy}}{\cal O} e^{\tau{\cal H}_{xy}}.
\end{align}
By using Eqs.~(\ref{supp_eq:7}) and~(\ref{supp_eq:8}), 
Green's functions in terms of the Majorana fermions $c_{lw}$ and $c_{lb}$ are written as
\begin{align}
 {\cal G}^{ww}_{l-l'}({\rm i}\nu_n)&=-\int_0^\beta d\tau\means{c_{lw}(\tau)c_{l'w}}_0 e^{{\rm i}\nu_n \tau}\nonumber\\
&=\frac{1}{{\cal L}}\sum_k \left[{\cal G}^0_k({\rm i}\nu_n)+\bar{\cal G}^0_k({\rm i}\nu_n)\right]e^{{\rm i}k(l-l')},\\
{\cal G}^{bb}_{l-l'}({\rm i}\nu_n)&=-\int_0^\beta d\tau\means{c_{lb}(\tau)c_{l'b}}_0 e^{{\rm i}\nu_n \tau}={\cal G}^{ww}_{ll'}({\rm i}\nu_n),\\
{\cal G}^{wb}_{l-l'}({\rm i}\nu_n)&=-\int_0^\beta d\tau\means{c_{lw}(\tau)c_{l'b}}_0 e^{{\rm i}\nu_n \tau}\nonumber\\
&=\frac{1}{{\cal L}}\sum_k \left[-{\cal G}^0_k({\rm i}\nu_n)+\bar{\cal G}^0_k({\rm i}\nu_n)\right]e^{{\rm i}\theta_k+{\rm i}k(l-l')},\\
{\cal G}^{bw}_{l-l'}({\rm i}\nu_n)&=-\int_0^\beta d\tau\means{c_{lb}(\tau)c_{l'w}}_0 e^{{\rm i}\nu_n \tau}=-{\cal G}^{bw}_{l'l}({\rm i}\nu_n).
\end{align}

In the Kitaev model on the hyperhoneycomb lattice, the lowest-order non-vanishing term appears in the fourth order with respect to ${\cal H}_z$ in Eq.~(\ref{supp_eq:14}). 
This is due to the fact that the smallest loop on the hyperhoneycomb lattice is the ten-site one, which include four $z$ bonds. 
The fourth-order contribution to the free energy is given by
\begin{align}
 F^{(4)}=-\frac{T}{4!}\int_0^\beta &d\tau_1\int_0^\beta d\tau_2\int_0^\beta d\tau_3\int_0^\beta d\tau_4\nonumber\\
&\times\mean{T_\tau {\cal H}_{z}(\tau_1){\cal H}_{z}(\tau_2){\cal H}_{z}(\tau_3){\cal H}_{z}(\tau_4)}_{0c}.
\label{supp_eq:F4}
\end{align}
A typical term in $F^{(4)}$ is written as
\begin{align}
 \sum_{p, q, s, \Delta l, \Delta l'}\tilde{J}_{\Delta l,\Delta l'}
&\eta_{(p,q,s)1}\eta_{(p+\Delta l,q,s)1}
\nonumber\\
&\times
\eta_{(p,q+\Delta l',s)2}\eta_{(p+\Delta l,q+\Delta l',s)2},\label{supp_eq:9}
\end{align}
where $r=[(p,q,s),\zeta]$ is the position of a $z$ bond in the unit cell at $p\bm{a}+q\bm{b}+s\bm{c}$ and a sublattice index $\zeta=1,2$ (there are two $z$ bonds in a unit cell). The summations over $\Delta l$ and $\Delta l'$ are taken over all integers. The coefficient $\tilde{J}_{\Delta l,\Delta l'}$ in Eq.~(\ref{supp_eq:9}) is given as
\begin{align}
 \tilde{J}_{\Delta l,\Delta l'}\propto J_z^4 T\sum_{i\nu_n}{\cal G}_{\Delta l}^{bb}(i\nu_n) {\cal G}_{\Delta l}^{ww}(i\nu_n){\cal G}_{\Delta l'}^{wb}(i\nu_n) {\cal G}_{\Delta l'}^{bw}(i\nu_n).
\end{align}
By taking the Matsubara sum, $\tilde{J}_{\Delta l,\Delta l'}$ is written as
\begin{align}
 \tilde{J}_{\Delta l,\Delta l'}=\frac{J_z^4}{J^3}\tilde{h}
 (T/J,\Delta l,\Delta l'),
\end{align}
where $\tilde{h}
(x,\Delta l,\Delta l')$ is a function independent of $J$ and $J_z$. 
Other terms in Eq.~(\ref{supp_eq:F4}) are also written in a similar form.
Hence, we finally obtain the relation
\begin{align}
F^{(4)}= \frac{J_z^4}{J^3}h(T/J, \{\eta_r\}),
\end{align}
where $h(x, \{\eta_r\})$ is a function independent of $J$ and $J_z$.
Thus, if there is a finite-$T$ phase transition, the critical temperature $T_c$ should be scaled as 
\begin{align}
 T_c\propto \frac{J_z^4}{J^3}.
\end{align}



\begin{thebibliography}{99}
\bibitem{fn}
The finite-size scalings for $C_v$ and $\tilde{{\cal W}}_c$ are not satisfactory for the present data sets, presumably because of the limited system sizes.
\end{thebibliography}
\end{document}